\documentstyle[12pt,epsfig,a4wide,rotate]{article}
\begin{document}
%
%
%
\def\Journal#1#2#3#4{{#1} {\bf #2}, #3 (#4)}
%
%
\def\NCA{\em Nuovo Cimento}
\def\NIM{\em Nucl. Instrum. Methods}
\def\NIMA{{\em Nucl. Instrum. Methods} A}
\def\NPB{{\em Nucl. Phys.} B}
\def\PLB{{\em Phys. Lett.}  B}
\def\PRL{\em Phys. Rev. Lett.}
\def\PRD{{\em Phys. Rev.} D}
\def\ZPC{{\em Z. Phys.} C}
\def\EPC{{\em Eur. Phys. J.} C}
\def\CMC{{\em Comp. Phys. Comm.} }
\newcommand{\tpm}        {\mbox{$\,\pm\,$}}
\newcommand{\ra}         {\mbox{$\rightarrow$}}
\newcommand{\GeVs}      {\mbox{${\,\rm GeV}^2$}}
\newcommand{\GeV}       {\mbox{${\,\rm GeV}$}}
\newcommand{\MeV}       {\mbox{${\,\rm MeV}$}}
\newcommand{\sqs}       {\mbox{$\sqrt{s}$}}
\newcommand{\qsq}       {\mbox{${Q^2}$}}
\newcommand{\sig}       {\mbox{$\sigma$}}
\newcommand{\sigb}      {\mbox{$\sigma\cdot B$}}
\newcommand{\gcc}       {\mbox{$\Gamma_{c\bar c}$}}
\newcommand{\gha}       {\mbox{$\Gamma_{had}$}}
\newcommand{\gch}       {\mbox{$\frac{\Gamma_{c\bar c}}{\Gamma_{had}}$}}
\newcommand{\gchfb}     {\mbox{$\gch \cdot f(c\ra D,\Lambda) \cdot B$}}
\newcommand{\zn}        {\mbox{$\rm{Z}^0$}}
\newcommand{\ccb}        {\mbox{$c\bar{c}$}}
\newcommand{\ds}         {\mbox{$D^{\ast}$}}
\newcommand{\dsp}        {\mbox{$D^{\ast +}$}}
\newcommand{\dspm}       {\mbox{$D^{\ast \pm}$}}
\newcommand{\dss}        {\mbox{$D_s$}}
\newcommand{\dssp}       {\mbox{$D_s^+$}}
\newcommand{\dsspm}      {\mbox{$D_s^{\pm}$}}
\newcommand{\dz}         {\mbox{$D^{0}$}}
\newcommand{\dc}         {\mbox{$D^+$}}
\newcommand{\lc}         {\mbox{$\Lambda_c$}}
\newcommand{\lcp}        {\mbox{$\Lambda_c^+$}}
\newcommand{\xic}        {\mbox{$\Xi_c$}}
\newcommand{\omc}        {\mbox{$\Omega_c$}}
\newcommand{\fcdz}       {\mbox{$f(c \rightarrow D^0)$}}
\newcommand{\fcdc}       {\mbox{$f(c \rightarrow D^+)$}}
\newcommand{\fcdss}      {\mbox{$f(c \rightarrow D_s^+)$}}
\newcommand{\fclc}       {\mbox{$f(c \rightarrow \Lambda_c^+)$}}
\newcommand{\fcds}       {\mbox{$f(c \rightarrow D^{\ast +})$}}
\newcommand{\fcdl}       {\mbox{$f(c \rightarrow D,\Lambda)$}}
\def\dskpi{ {\dsp}~\rightarrow~\dz~\pi^{+}_{s}%
        \rightarrow~(K^{-}~\pi^{+})~\pi^{+}_{s} }
\def\dssphipi{ {\dssp}~\rightarrow~\phi~\pi^{+}%
        \rightarrow~(K^{-}~K^{+})~\pi^{+} }
\def\dzkpi{ {\dz}~\rightarrow~K^{-}~\pi^{+} }
\def\dpkpi{ {\dz}~\rightarrow~K^{-}~\pi^{+} }
\def\dckpipi{ {\dc}~\rightarrow~K^{-}~\pi^{+}~\pi^{+} }
\def\lcpkpi{ {\lcp}~\rightarrow~p~K^{-}~\pi^{+} }
%
 
\begin{titlepage}
 
\title{\Large{\centerline{\bf Charm Hadron Production Fractions}}}

\author{Leonid Gladilin
~\footnote{gladilin@mail.desy.de}
~\footnote{On leave from Moscow State University, supported by GIF, contract I-0444-176.07/95} \\
\it{Inst. of Experimental Physics II, University of Hamburg}}
\maketitle
 
\begin{abstract}

The world average values for the probabilities that a charm quark fragments
into \dsp~ and \dssp~ have been calculated to be
$\fcds~=~0.235\tpm0.007~(\pm0.007)$ and $\fcdss~=~0.101\tpm0.009~(\pm0.025)$,
respectively.
The average values for \dz, \dc~ and \lcp~ have been also calculated.

\end{abstract}
 
 
\setcounter{page}{1}
\thispagestyle{empty} 

\end{titlepage}
 

\section*{\bf Introduction}
\label{sec:intro}

The probabilities that a charm quark fragments into \dsp, \dssp~ and
other charm hadrons are not calculable in perturbative QCD.
To obtain them one can use data
on charm production in $e^+e^-$ annihilations.
The most comprehensive charm measurements
were performed by
the CLEO~\cite{cleo8889,cleo91}
and ARGUS~\cite{argus88,argus91,argus92} collaborations
at centre-of-mass energies of about 10\GeV~
and by
the OPAL~\cite{opal96,opal98},
ALEPH~\cite{aleph99}
and DELPHI~\cite{delphi98,delphi99} collaborations
in \zn~ decays. We will use their measurements performed
with charm hadrons reconstructed in the following decay
modes\footnote{The charge conjugated processes are also included.}:
\begin{equation}
   \dskpi,
   \label{dskpi}
\end{equation}
\begin{equation}
   \dssphipi,
   \label{dssphipi}
\end{equation}
\begin{equation}
   \dzkpi,
   \label{dzkpi}
\end{equation}
\begin{equation}
   \dckpipi,
   \label{dckpipi}
\end{equation}
\begin{equation}
   \lcpkpi,
   \label{lcpkpi}
\end{equation}

The Particle Data Group (PDG) branching ratios for the reference decay modes are~\cite{pdg98}:
$B(\dsp \ra~ \dz\pi^+)~=~0.683\tpm0.014$,
$B(\dssp \ra~ \phi\pi^+)~=~0.036\tpm0.009$,\\
$B(\dz \ra~ K^-\pi^+)~=~0.0385\tpm0.0009$,
$B(\dc \ra~ K^-\pi^+\pi^+)~=~0.090\tpm0.006$ ~~and\\
$B(\lcp \ra~ p K^-\pi^+)~=~0.050\tpm0.013$.
 

\section*{\bf CLEO and ARGUS Measurements}
\label{sec:clar}

CLEO and ARGUS measurements are summarized in 
table~1.
In the case if two errors are quoted then the first error is statistical
and the second is systematic.
The \dssp~ production cross sections are corrected for the
branching ratio B($\phi \rightarrow K^{-} K^{+}$) = 0.491\tpm0.008~\cite{pdg98}.

\begin{table}[hbpt]
\begin{center}

\begin{tabular}{|c|c|c|} \hline

Particle   & CLEO & ARGUS \\
           & \sigb~ (pb) & \sigb~ (pb) \\
\hline
\hline
\dsp
& 17.0 \tpm 1.5 \tpm 1.4 & 14.1 \tpm 1.5 \tpm 1.4 \\
\hline
\dssp
& 7.2 \tpm 1.9 \tpm 1.0 & 7.5 \tpm 0.8 \tpm 0.7 \\
\hline
\dz
& 52 \tpm 5 \tpm 4 & 43.8 \tpm 5.6 \\
\hline
\dc
& 51 \tpm 7 \tpm 2 & 50.0 \tpm 6.9 \\
\hline
\lcp
& 10.0 \tpm 1.5 \tpm 1.5 & 9.0 \tpm 1.2 \tpm 1.0 \\
\hline

\end{tabular}

\label{t:clar}
\caption{
{Measured cross sections times branching ratios, \sigb, for the production
of the charm hadrons detected through the reference decay modes at
centre-of-mass energies of about 10\GeV.}}
\end{center}
\end{table}

The total hadronic cross section at 10.55\GeV~ was measured
to be\\
3.33\tpm0.05\tpm0.21~nb~\cite{cleo84}.
In Ref.~~\cite{cleo8889}
the fraction of \ccb~ in the total hadronic cross section
was estimated with a Monte Carlo calculation to be 0.37\tpm0.02 (syst.)
and thus the total cross section into charm particles was calculated
to be
\begin{equation}
{\sigma(c) = 2.46\tpm0.04(stat.)\tpm0.20(syst.)\,nb ~.}
   \label{cc1055}
\end{equation}

Dividing the measured values from table~1 by the total charm cross
section and the branching ratio for the reconstruction mode,
charm hadron production fractions have been obtained.


\section*{\bf LEP Measurements}
\label{sec:lep}

LEP measurements of the production rates of \dz, \dc, \dsp~ and \lcp~ hadrons
in \zn~ decays are summarized in
table~2.
The OPAL collaboration combined
the decay channel $\dssp~ \ra~ \bar{K}^{* 0} K^+$ and the channel~(2) with
the ratio fixed to the PDG ratio of branching fractions~\cite{pdg98}:
B($\dssp~ \ra~ \bar{K}^{* 0}K^+$)/B($\dssp~ \ra~ \phi\pi^+$) = 0.95\tpm0.10.
The result is expressed in terms of
B($\dssp~ \ra~ \phi\pi^+$) and
the uncertainty in this ratio is treated as a systematic error.
All \dssp~ product branching ratios in
table~2
are corrected for the
branching ratio B($\phi \rightarrow K^{-} K^{+}$).

\begin{table}[hbpt]
\begin{center}

\begin{tabular}{|c|c|c|c|} \hline

Particle   & OPAL & ALEPH & DELPHI \\
           & \gchfb~ & \gchfb~ & \gchfb~ \\
           & (\%) & (\%) & (\%) \\
\hline
\hline
\dssp
& 0.056 \tpm 0.015 \tpm 0.007 & 0.072 \tpm 0.012 \tpm 0.004 & 0.076 \tpm 0.007 \tpm 0.007 \\
\hline
\dz
& 0.389 \tpm 0.027 $^{+0.026}_{-0.024}$ & 0.370 \tpm 0.011 \tpm 0.023 & 0.360 \tpm 0.010 \tpm 0.021 \\
\hline
\dc
& 0.358 \tpm 0.046 $^{+0.025}_{-0.031}$ & 0.368 \tpm 0.012 \tpm 0.020 & 0.349 \tpm 0.012 \tpm 0.021 \\
\hline
\lcp
& 0.041 \tpm 0.019 \tpm 0.007 & 0.067 \tpm 0.007 \tpm 0.004 & 0.074 \tpm 0.015 \tpm 0.009 \\
\hline

\end{tabular}

\label{t:lep}
\caption{
{LEP measurements of the products of the partial decay width of the \zn~
into \ccb~ quark pairs, \gch, charm hadron production fractions, \fcdl,
and corresponding branching ratios.}}
\end{center}
\end{table}

Charm hadron production fractions can be obtained dividing the measured
products by the branching ratios and the Standard Model value
of \gch~\cite{gch_theor}:
\begin{equation}
{\gch~ = 0.1719\tpm0.0017 ~.}
   \label{gch_theor}
\end{equation}

There ~are ~four ~LEP ~measurements ~which ~can ~be ~used ~for ~the ~\fcds~~ calculation.
The ~OPAL ~collaboration ~has ~done ~a double ~tagged ~measurement of
$\fcds\cdot B(\dsp\ra\dz\pi^+)$~\cite{opal96}. Using $B(\dsp\ra\dz\pi^+) = 0.683$
\cite{pdg98} they have obtained:
\begin{equation}
{\fcds~ = 0.222\tpm0.014\tpm0.014 ~.}
   \label{fcds_opal}
\end{equation}

Another double tagged measurement has been done by the DELPHI
collaboration~\cite{delphi99}:
\begin{equation}
{\fcds\cdot B(\dsp\ra\dz\pi^+) = 0.174\tpm0.010\tpm0.004 ~,}
   \label{fb_delphi}
\end{equation}
\begin{equation}
{\fcds~ = 0.255\tpm0.015\tpm0.006 ~.}
   \label{fcds_delphi}
\end{equation}

The ALEPH collaboration has done \dspm~ rate measurement~\cite{aleph99}.
Using $B(\dsp\ra\dz\pi^+)$ and $B(\dz\ra K^-\pi^+)$ values from \cite{pdg98}
they have obtained:
\begin{equation}
{\fcds~ = 0.2333\tpm0.0102\tpm0.0084 ~.}
   \label{fcds_aleph}
\end{equation}

Another \dspm~ rate measurement has been done by the OPAL
collaboration~\cite{opal96}:
\begin{equation}
{\gch\cdot\fcds\cdot B(\dsp\ra\dz\pi^+)\cdot B(\dz\ra K^-\pi^+) = (1.041\tpm0.020\tpm0.040)\times {10}^{-3} ~.}
   \label{rfb_opal}
\end{equation}

Using PDG branching ratios and the Standard Model value for \gch~, the result
can be transformed to \fcds~ value:
\begin{equation}
{\fcds~ = 0.2303\tpm0.0044\tpm0.0091 ~.}
   \label{fcds_opal2}
\end{equation}

Two OPAL's \fcds~ measurements are statistically and systematically correlated.
To calculate the correlations one can use the result for \gch~ obtained
by the OPAL collaboration~\cite{opal96} from the
measurements~(\ref{fcds_opal}) and~(\ref{rfb_opal}):
\begin{equation}
{\gch~ = 0.180\tpm0.011\tpm0.012 ~.}
   \label{gch_opal}
\end{equation}

Note that old value
$B(\dz \ra~ K^-\pi^+)~=~0.0383$~\cite{pdg96} has been used by the OPAL
collaboration for the \gch~ calculation.


\section*{\bf Average Charm Hadron Rates}
\label{sec:average}

Tables~3 and~4
contain all charm hadron production
fractions discussed above and their average values.
In the case if two errors are quoted then the first error is statistical
and the second is systematic.
The errors in parentheses for the average values are due to the uncertainties
in the charm hadron branching ratios.
They can be ignored in the case if the values are used for comparisons
to measurements done with the same decay mode and the same
branching ratio for the mode.

\begin{table}[hbt]
\begin{center}
\begin{tabular}{|c|c|c|c|} \hline

   & \fcdz ~(\%) & \fcdc ~(\%) & \fclc ~(\%) \\
\hline
\hline
CLEO
& 54.9 \tpm 5.4 \tpm 6.1 & 23.0 \tpm 3.2 \tpm 2.1 & 8.1 \tpm 1.2 \tpm 1.4\\
\hline
ARGUS
& 46.2 \tpm 7.0 & 22.6 \tpm 3.6 & 7.3 \tpm 1.0 \tpm 1.0\\
\hline
ALEPH
& 55.9 \tpm 1.7 \tpm 3.5 & 23.8 \tpm 0.8 \tpm 1.3 & 7.8 \tpm 0.8 \tpm 0.4\\
\hline
DELPHI
& 54.4 \tpm 1.5 \tpm 3.2 & 22.6 \tpm 0.8 \tpm 1.4 & 8.6 \tpm 1.8 \tpm 1.0\\
\hline
OPAL
& 58.8 \tpm 4.1 $^{+4.0}_{-3.7}$ & 23.1 \tpm 3.0 $^{+1.6}_{-2.0}$ & 4.8 \tpm 2.2 \tpm 0.8\\
\hline
\hline
Average
& 54.9 \tpm 2.3 (\tpm1.3) & 23.2 \tpm 1.0 (\tpm1.5) & 7.6 \tpm 0.7 (\tpm2.0)\\
\hline
\end{tabular}

\label{t:ctod3}
\caption{Measured and average probabilities that a charm quark fragments
into \dz, \dc~ and \lcp.
The average probabilitity errors in parentheses are due to the uncertainties
in the charm hadron branching ratios.}
\end{center}
\end{table}
\begin{table}[hbt]
\begin{center}
\begin{tabular}{|c|c|c|} \hline

   & \fcdss ~(\%) & \fcds ~(\%) \\
\hline
\hline
CLEO
& 8.1 \tpm 2.2 \tpm 1.3 & 26.3 \tpm 2.4 \tpm 3.0 \\
\hline
ARGUS
& 8.5 \tpm 0.9 \tpm 1.0 & 21.8 \tpm 2.3 \tpm 2.8 \\
\hline
ALEPH
& 11.5 \tpm 1.9 \tpm 0.7 & 23.3 \tpm 1.0 \tpm 0.8 \\
\hline
DELPHI
& 12.4 \tpm 1.1 \tpm 1.2 & 25.5 \tpm 1.5 \tpm 0.6 \\
\hline
OPAL
& 9.0 \tpm 2.4 \tpm 1.1 & 22.2 \tpm 1.4 \tpm 1.4 \\
&                       & 23.0 \tpm 0.4 \tpm 0.9 \\
\hline
\hline
Average
& 10.1 \tpm 0.9 (\tpm2.5) & 23.5 \tpm 0.7 (\tpm0.7) \\
\hline
\end{tabular}

\label{t:ctod2}
\caption{Measured and average probabilities that a charm quark fragments
into \dssp~ and \dsp.
The average probabilitity errors in parentheses are due to the uncertainties
in the charm hadron branching ratios.}
\end{center}
\end{table}

To average the measurements a standard weighted least-squares procedure
has been used~\cite{pdg98}. In the case if statistical and systematic errors
were quoted separately they have been added in quadrature and the combined
error has been used. The OPAL collaboration quoted asymmetric systematic
errors for \dz~ and \dc~ measurements. The errors with larger absolute
values have been used for averages in these two cases.

The average takes into account:
\begin{itemize}
\item
{common ARGUS/CLEO errors due to the errors in the total charm cross
sections~(\ref{cc1055});}
\item
{common LEP errors due to the error in the Standard Model value
of \gch(\ref{gch_theor});}
\item
{correlations between two OPAL's \fcds~ measurements.}
\end{itemize}
Other experimental errors were assumed to be uncorrelated.

A sum of the average \dz, \dc, \dssp~ and \lcp~ production
fractions equals to $95.8\%$ leaving a room for the fragmentation
of a charm quark to \xic~ and \omc.

\section*{\bf Acknowledgments}
\label{sec:ackno}

I thank Erich Lohrmann, Uri Karshon and Olaf Deppe
for useful discussions and help in the work with
the bibliography.


\end{document}